\documentclass[prl,twocolumn,groupedaddress]{revtex4}
\usepackage{graphics,epsfig}
\usepackage{amsmath,amssymb}

\begin{document}

\title{Evolution Management in a Complex Adaptive System: Engineering the
Future}
\author{David M.D. Smith$^1$ 
and Neil F. Johnson$^2$}
\address{$^1$ Mathematics Department, Oxford University,  Oxford, OX1 2EL,
U.K., \\
$^2$ Physics Department, Oxford University, Oxford OX1 3PU, U.K.}
\date{\today}

\begin{abstract} 
We examine the feasibility of predicting and subsequently 
managing the future evolution
of a Complex Adaptive System. Our archetypal system 
mimics a competitive population of mechanical, biological, informational 
or human objects. We show that short-term prediction yields corridors along which the
system will, with very high probability, evolve. We then show how small
amounts of `population engineering' can be undertaken in order  to steer  the
system away from any undesired regimes which have been predicted. Despite the system's many degrees
of freedom and inherent stochasticity, this dynamical `soft' control over 
future risk requires
only minimal knowledge about the population's composition. 

\end{abstract} 

\maketitle

Complex Adaptive Systems (CAS) are of great interest to theoretical physicists
because they comprise large numbers of interacting components or `agents'
which, unlike particles in traditional physics, may change their behavior based
on past experience \cite{bocc}. Such adaptation yields complicated feedback
processes at the microscopic level, which in turn generate complicated global
dynamics at the macroscopic level.   CAS also arguably represent the `hard'
problem in  biology, engineering, computation and sociology
\cite{bocc}. Depending on the application domain, the agents in CAS may 
represent species, people, bacteria, cells, computer hardware or software, and
are typically fairly numerous, e.g. $10^2-10^3$
\cite{bocc,1}.

There is also great practical interest in the problem of predicting and subsequently 
controlling a Complex Adaptive System.
Consider the enormous task facing 
a Complex Adaptive System `manager' in charge of overseeing some complicated 
computational, biological, medical, sociological or even economic system. He 
would certainly like to be able to predict its future evolution 
with sufficient accuracy that he could  
foresee the system heading towards any `dangerous' areas. However, prediction is not enough.
He also needs to be able to steer the system away from this 
dangerous regime. Furthermore, the CAS-manager needs to be able to achieve this {\em without} detailed 
knowledge of the present state of its thousand different components, nor does he want to
have to shut down the system completely. Instead he is seeking some form of `soft' 
control. Unfortunately, his task looks hopeless.
Even in purely deterministic systems
with only a few degrees of freedom, it is well known that highly complex
dynamics such as chaos can arise \cite{strogatz} making both prediction and 
control very difficult -- for example, the `butterfly effect' wherein small
perturbations  have huge unpredictable  consequences. Consequently, one would
think that things would be considerably worse in  a CAS, given the much larger
number of interacting objects.   As an additional complication, a CAS may also
contain stochastic processes at the microscopic and/or macroscopic levels,
thereby adding an inherently random element to the system's dynamical
evolution.  The Central Limit Theorem tells us that the combined effect of a
large number of stochastic processes tends fairly rapidly to a Gaussian
distribution. Hence, one would guess that even with reasonably complete
knowledge of the present and past states of the system, the evolution would be
essentially diffusive and hence difficult to control without imposing
substantial global constraints.

In this paper, we examine this question of evolution management for a
simplified, yet highly  non-trivial model of a CAS. We show that a surprising
level of prediction and subsequent control are indeed possible. 
First we show that with very little knowledge about the system's past
behavior, one can produce corridors (Future-Casts) along which the system will
subsequently move, characterized by their width (Characteristic Stochasticity) 
and their average direction (Characteristic Direction). Although these 
corridors evolve as the system evolves,
 at any particular point in time  
they provide an accurate prediction regarding the subsequent evolution of the 
system.
We then show that if the Future-Cast predicts significant future risk, the system's 
subsequent evolution can be steered to a safer regime 
via `population engineering', i.e. by
introducing small perturbations to the  population's heterogeneity. 
Despite the many degrees of freedom and inherent
stochasticity both at the microscopic and macroscopic levels, this global
control  requires only minimal knowledge and intervention on the part of a CAS
manager.  For the somewhat simpler case of Cellular Automata, Israeli and
Goldenfeld
\cite{golden} have recently obtained the remarkable result that computationally
irreducible physical processes can become computationally reducible at a
coarse-grained level of description. Based on our findings, one could speculate
that similar ideas hold for  populations of decision-taking, adaptive
agents. We finish the paper by discussing a number of possible practical applications of 
our findings.

It is widely believed (see for example, Ref. \cite{casti1}) that Arthur's
El Farol Bar Problem \cite{2} provides a representative toy model for
CAS's which comprise a population of objects competing for some limited global
resource  (e.g. space in an overcrowded area).  To make this model more
complete in terms of real-world complex systems,  the effect of network
interconnections has recently been incorporated \cite{nets}.  As mentioned
later, our present analysis also applies to such networked populations.  The El
Farol Bar Problem concerns the collective decision-making of a group of
potential bar-goers (i.e. agents) who use limited global information to predict
whether they should attend a potentially overcrowded bar on a given night each
week. The Statistical Mechanics community has adopted a binary version of this
problem, the so-called Minority Game (MG)
\cite{3,6}, as a new form of Ising model which is worthy of study in its own
right because of its highly non-trivial dynamics.  Here we consider a
generalized version of such multi-agent binary games which (a) incorporates a
finite time-horizon $T$ over which agents remember their strategies' past
successes, to reflect the fact that the more recent past should have more
influence than the distant past, (b)  allows for  fluctuations in agent
numbers, since agents might  only participate if they possess a strategy with a
sufficiently high success rate, and (c) allows  for a general reward structure
thereby disposing of the MG's restriction to automatically rewarding the
minority \cite{6}.  The formalism is applicable to any CAS which can be mapped
onto a population of $N$ objects which repeatedly taking actions in
some form of global `game'. For simplicity, we restrict ourselves here to
simply invoking competition for  a limited resource $L$. Our model therefore
incorporates  the features typically   associated with complex systems: strong
feedback, adaptation, interconnectivity etc.  At each timestep $t$, each agent
makes a (binary) decision
$a_{\mu(t)}$ in response to some global information $\mu(t)$.  This global
information is a bitstring of length $m$, and may for example represent the
history of past global outcomes. The global outcome at a given timestep 
is  based on the
aggregate action of the agents and the value of the global resource level $L$.
Each agent holds
$k$ strategies (comprising a response to every possible history) employing the one which would have proved most successful over the last $T$ timesteps. By assigning
these randomly to each agent, we mimic the effect of large-scale heterogeneity
in the population. The strategy allocation is fixed at the start of the game,
and can be described by a tensor of rank $k$ or `Quenched Disorder Matrix'
(QDM) \cite{6}. Adding network connections simply  has the effect of
redistributing elements within  the QDM.  The agents' aggregate action at each
timestep $t$ is represented by
$D(t)$, and $S(t)=S(t-1)+D(t-1)$ gives the current global output value
\cite{note}.  Stochasticity arises via coin-tosses at both the microscopic
level (to resolve an agent's tied strategies) and the macroscopic level (to
resolve any ties when deciding the global outcome). This  stochasticity implies
that for a given QDM, the system's output  is not unique. In short, 
the future evolution
of the system results from the  {\em time-dependent} interplay of
{\em time-dependent} deterministic and stochastic  processes. We refer to the
set of all possible future trajectories  of the game's output at $t\geq 1$  
timesteps in
the future, as the {\em Future-Cast} distribution.

The game's dynamics can be transferred into a time-horizon space
$\underline{\Gamma_{t}}$ spanned by all  possible combinations of the last
$m+T$ global outcomes (or equivalently, the winning actions)
 \cite{6}. For a binary game, $\underline{\Gamma_{t}}$ has dimension
$2^{m+T}$. For any given time-horizon state 
$\Gamma_{t}$ in this space, there exists a unique score vector
$\underline{G(t)}$ whose element $G_{R}(t)$ is the score for strategy $R$ at
time $t$. 
Each time a particular time-horizon state is reached, the 
actions of the agents holding
strategies whose scores are not tied, or agents holding tied strategies which 
prescribe the same action, will necessarily be the same. 
In addition, the number of remaining agents (i.e. those holding tied
strategies prescribing different actions, which need to be resolved 
via a coin-toss) will also be the same. 
Subsequently, the
probability distribution of $D(t)$ will be identical each 
time this time-horizon
state occurs.
The probabilities associated with the global outcomes which represent the 
transitions between these time-horizon states are also static. Hence it is 
possible to construct a Markov
Chain description for the evolution of the probabilities
$\underline{P(\Gamma_{t})}$ for these time-horizon states:
\begin{equation}
  \underline{P(\Gamma_{t})} ~= 
  ~\underline{\underline{T}}
  ~\underline{P(\Gamma_{t-1})}.\end{equation}  
The transition matrix
$\underline{\underline{T}}$ is time-independent and sparse since there are only
two possible global outcomes for each state. The number of non-zero elements
in the matrix is thus
$\le2^{(m+T+1)}$. These values can be generated directly from the QDM \cite{9}. 
It is straightforward to obtain the stationary state solution
of Eq. (1) in order to calculate the system's time-averaged macrosopic
quantities. Generating the {\em Future-Cast} probability distributions involves  mapping from the internal (time-horizon) state dynamics of the system to its global output.
This requires
(i) the probability distribution of
$D(t)$ for a given
  time-horizon $T$, (ii) the corresponding global outcome for
 a given $D(t)$, and (iii) an output generating algorithm expressed in terms of
$D(t)$. We know that in the transition matrix, the probabilities represent the
summation over a distribution which is binomial in the case where the agents
are limited to two possible decisions. Using the output generating algorithm,
we can construct an adjacency matrix
$\underline{\underline{\Upsilon}}$ to the transition matrix
$\underline{\underline{T}}$, with the same dimensions. The elements of
$\underline{\underline{\Upsilon}}$ contain probability functions
corresponding to the non-zero  elements of the transition matrix, together with
the discrete convolution operator. For the incremental algorithm 
described above, we define and use a convolution operator
$\otimes$  such that $(f\otimes g)\mid_{i} ~= \sum_{j=-\infty}^{\infty} 
f(i-j)\times g(j)$ (see Ref. \cite{9} for full mathematical details). 
Consider an arbitrary timestep in the game, and 
label it as
$t=0$ for convenience. The adjacency matrix can then be applied to a vector
$\underline{\varsigma(S,t=0)}$, where the element of
$\underline{\varsigma(S,0)}$ corresponding to the current time-horizon state
comprises a probability distribution function for the current output value.
Since $\underline{\varsigma(S,t)}~=~
\underline{\underline{\Upsilon}}^{t}
\underline{\varsigma(S,0)}$,  the Future-Cast at $t$ timesteps in the future,
$\Pi(S,t)$, is given by:
\begin{equation}
\Pi(S,t) = \sum_{i=1}^{2^{(m+T)}}\varsigma_i(S,t).\end{equation}  Due to the
state dependence of the Markov Chain, this Future-Cast probability distribution
$\Pi$ is non-Gaussian. Consider $t=1$. Since we are not interested in
transients, we really need  a `steady state' form $\Pi_1 =
\big<\Pi(S,1)\big>_\infty$ representing a time-average over an infinitely long
period.  Fortunately, we have the steady state solutions of
$\underline{P(\Gamma)} ~=\underline{\underline{T}}~\underline{P(\Gamma)} $
which are the (static) probabilities of being in a given state at any time. By
representing these probabilities as the appropriate functions, we can construct
an initial vector 
$\underline{\kappa}$, which is equivalent to
$\underline{\varsigma(S,0)}$  \cite{9}. 
Hence we can generate the {\em Characteristic
Future-Cast} $\Pi_1$, describing the characteristic behavior of the Future-Cast
projected forward from a general time $t$, for a given QDM. The element
$\kappa_{i}$ is simply the point $(0,P_{i}(\Gamma))$. Characteristic
Future-Casts for any number of timesteps
$t\geq 1$ into the future, can be generated by simply 
premultiplying $\underline{\kappa}$ by
$\underline{\underline{\Upsilon}}^t$: i.e. use Eq. (2) with 
$\underline{\varsigma}~=~\underline{\underline{\Upsilon}}^t~\underline{\kappa}$.  
Hence the Characteristic Future-Cast over
$t$ timesteps is simply the Future-Cast of length
$t$ from all the $2^{m+T}$ possible initial states, with each contribution
being given the appropriate weighting factor.  Note that $\Pi_{t}$ is {\em not}
equivalent to the convolution of
$\Pi_{1}$ with itself $t$ times, and hence is not necessarily Gaussian. In
other words, the Central Limit Theorem does {\em not} provide a good estimate
of the future behavior of the system. The system is only quasi-diffusive {\em
at best}.


\begin{figure}[ht]
\includegraphics[width=3.6in]{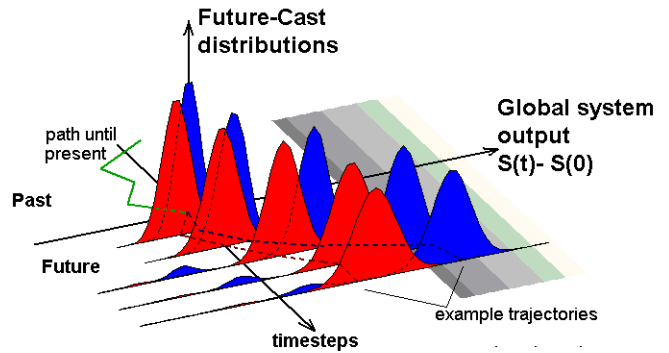}
\vskip0.1in
\caption{Evolution of the unperturbed Future-Cast probability distribution
(blue solid) during a typical run of a game with the Quenched Disorder Matrix 
(QDM) of Fig. 2.  
The region of large positive deviation
$S(t)$ is considered dangerous territory.  Red distributions show corresponding
evolution following a minor QDM perturbation (i.e. population engineering of
Fig. 2) introduced at $t=0$. Examples of future trajectories also shown. Note that $t=0$ labels an arbitary timestep chosen
after initial transients have disappeared.} \label{fig:figure1}
\end{figure}


Figure 1 shows a typical example of the evolution of the Future-Cast at $t$
timesteps ahead of the present timestep (which we label $t=0$). The Future-Cast
acts to provide non-Gaussian `corridors' along which the system subsequently
evolves. The reason for the non-diffusive behavior is that, unlike the standard
binomial paths set up during a simple coin-toss experiment, not all paths are
realized at every timestep. The stochasticity generated at a given timestep,
and hence the possible future paths, are {\em conditional} on the system's past
history.  Now suppose that an external CAS manager decides it  dangerous for
the system to have a  large positive $S(t)$ for
$t>0$. Figure 2 shows the corresponding QDM for
$t<0$, together with the QDM {\em perturbation} which the system manager decides
to introduce at 
$t = 0$.  Such `population engineering' can be achieved by
switching on/off, rewiring or 
reprogramming a group of agents in a situation where the agents
are accessible objects, or by introducing some form of communication channel -- 
or even a more evolutionary approach whereby a small subset of agents 
(`species') 
are removed
from the population and a new subset added to replace them. This 
evolutionary mechanism need neither be completely deterministic
(i.e. knowing exactly how the form of the QDM changes) nor completely random
(i.e. a random perturbation to the QDM). In this sense, it seems quite close to
some modern ideas of biological evolution, whereby there is some purpose mixed
with some randomness.  Figure 1 shows the impact that this relatively minor
perturbation has on the Future-Cast. In particular, the system gets steered
`away from danger' (i.e. toward smaller $S(t)$ values). Note that a 
substantial reduction 
in future risk has been 
achieved {\em without} needing to know the microscopic details of each agent's
individual strategies, since each QDM corresponds to a macrostate in the
physical sense: i.e. it is only the {\em aggregate} 
number of agents holding each
strategy pair which matters, not what an individual agent is holding.

\begin{figure}[ht]
\includegraphics[width=3.6in]{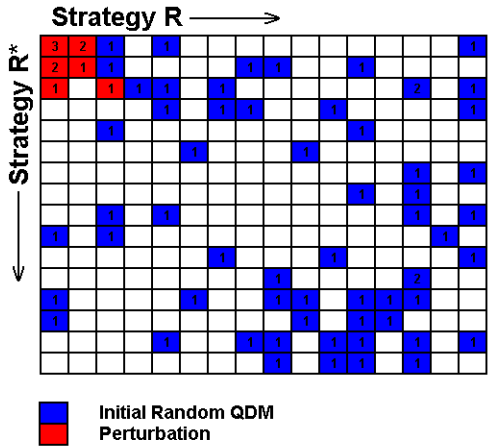}
\vskip0.1in
\caption{The unperturbed Quenched Disorder Matrix QDM (blue) 
and its perturbation (red)
used to generate Fig. 1. The i-j coordinates represent strategy
labels for the $k=2$ strategies per agent, $R$ and $R^*$. These binary strategies
$R$ and $R^*$ are 
ordered according to their decimal equivalent. Here 
$m=2$. The value in each box
represents the number of agents assigned 
that particular pair of strategies during the initial random allocation. 
An empty
bin implies no agent holds that particular strategy pair.}
\label{fig:figure2}
\end{figure}

Engineering an appropriate QDM perturbation involves understanding the
interplay between the (i) the mean of the Future-Cast distribution, referred to
as the {\em Characteristic Direction} which acts as a `drift' in terms of the
future output signal,  and (ii) the spread in the Future-Cast distribution, 
referred
to as the {\em Characteristic Stochasticity} which acts as `noise' in terms of
the future output signal. Figure 3 shows how these quantities vary for different QDMs
for the illustrative case of $m=1$ with a small population. This indicates
the effects of adding such a population as a perturbation to an unbiased
system.  In order to reliably steer $S(t)$ toward larger/smaller values, 
the  Characteristic
Direction must be much larger than the Characteristic Stochasticity. As shown,
the perturbation must therefore be biased toward the upper-left/lower-right
half of the QDM (but not both). This means that the perturbed population is
{\em less} adaptive than the unperturbed one (i.e. more agents hold two
identical strategies) {\em and} less heterogenous (i.e. more agents populate
the same region of the QDM). 
This observation explains why the QDM perturbation of Fig. 2 had the desired 
steering effect shown in Fig. 1.  By contrast for perturbations which are unbiased
in terms of the upper-left/lower-right half of the QDM, the Characteristic
Direction is zero and hence there is no net steering, while the Characteristic
Stochasticity is now large. These effects can be understood in terms of 
Crowd-Anticrowd formation in the strategy space \cite{3,6}.

\begin{figure}[ht]
\includegraphics[width=3.6in]{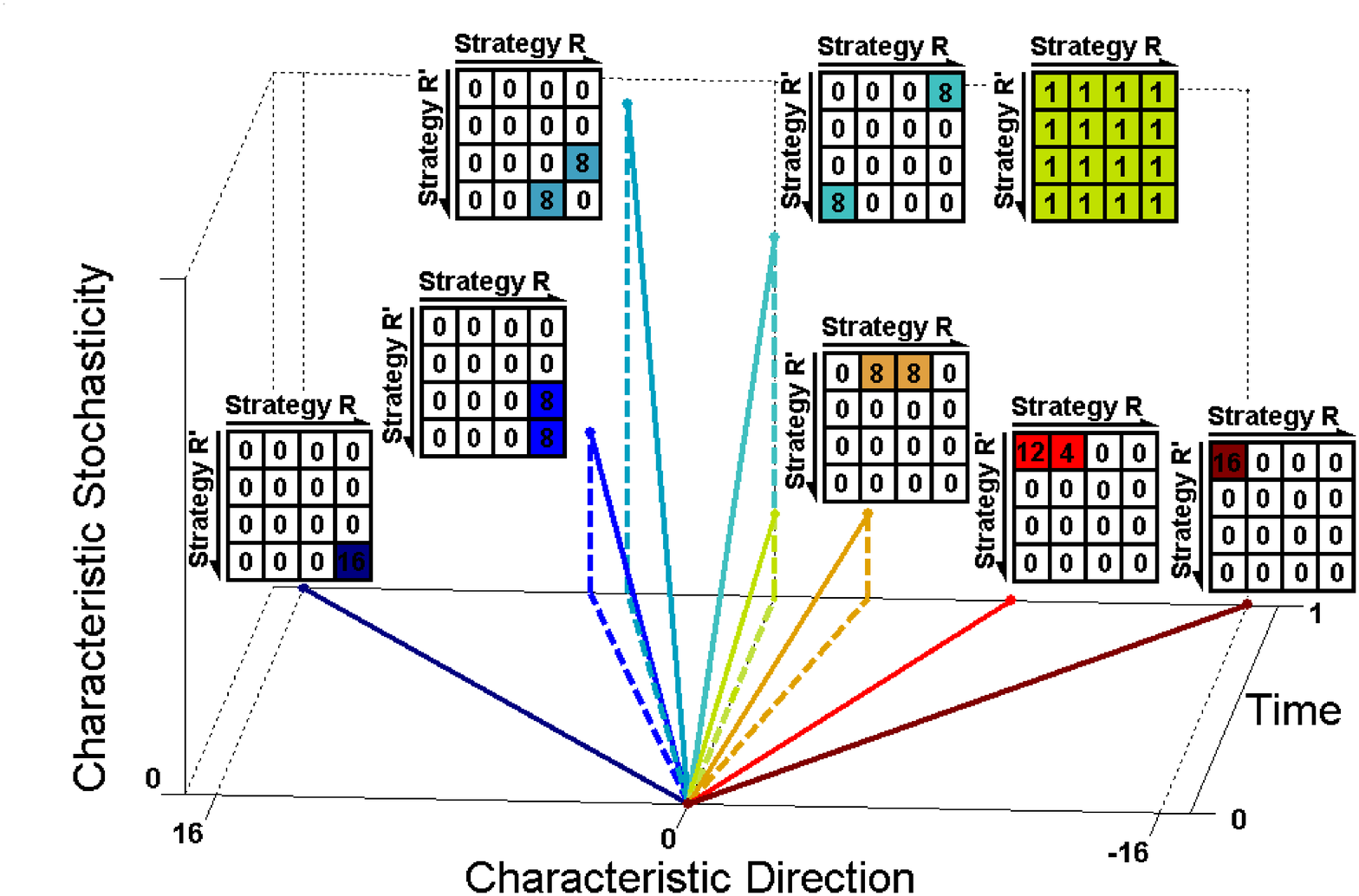}
\vskip0.1in
\caption{Characteristic Direction and Characteristic Stochasticity, for 
illustrative
QDMs with $m=1$ and $k=2$. Results are shown for $t=1$  timesteps
into the future.}
\label{fig:figure3}
\end{figure}

Finally, we give some examples to justify why we think our 
Complex-Adaptive-Systems control problem is so generic. Next-generation
aircraft wings may  contain thousands of autonomous mini-flaps placed along
the rear of a wing \cite{ilan}. Denoting  the binary actions of each miniflap as `up' and
`down', and rewarding flaps for their actions given the `resource level' $L$
(e.g. the plane's current tilt), Fig. 3 shows that one can simply
switch on a small number of additional miniflaps in order that the aircraft
then moves autonomously in a given direction. This is achieved 
{\em without} requiring sophisticated control of individual miniflaps, or 
inter-miniflap
communication  \cite{ilan}. In human health, there is a possible application in so-called 
dynamic
diseases. For example, Epilepsy is a dynamic disease involving 
sudden changes in the activity of millions of neurons. Our work raises hopes
that one could develop a relatively non-intrusive `brain defibrillator' 
using brief electrical stimuli
over a small part of the brain, rather than intrusive control over each and every one of
the constituent agents (i.e. neurons). In the area of cancer therapy, the tumor to be 
eradicated comprises a population of cancerous
and normal cells which compete for a limited resource (i.e. oxygen in blood supply, and space to grow).
It is possible that 
by understanding how the overall tumor cell population behaves, one could do some population
engineering of a small group of the malignant cells in order to steer the tumour toward benign 
status. Even in the immune system, where the body supposedly self-regulates itself
as a result of the interaction of hundreds of different biological processes (agents),
and where the corresponding  `steering wheel' remains unknown, our work suggests that one might be
able to engineer one part of the system so that it boosts or suppresses the overall 
immunological activity level. 
In a financial setting, where intervention in a market costs money, 
one could imagine that an external regulator 
could use our analysis to steer a particular market indicator or exchange rate 
into a desired range without 
having to invest huge amounts of money. Further details of these applications will be published 
elsewhere. In short, 
we believe that the present problem  
lies at the heart of complex
systems science both in terms of fundamental non-linear dynamical behavior and the consequences 
for
practical
safety management. 

\vskip0.2in

\noindent DMDS thanks Royal Bank of Scotland for funding under an EPSRC Case
studentship.  We are grateful to Prof. Ilan Kroo and Dr. David Wolpert  for
suggesting the aerospace application.


\begin{thebibliography}{99}
\bibitem{note}
  This approach can be used to generate equivalent results for any of the
  system's outputs (e.g. number of agents participating).
\end{thebibliography}
\end{document}